\documentclass[12pt]{article}
\usepackage{epsfig}
\usepackage{epsf}
\usepackage{bm}
\begin{document}

\vspace{2cm}

{\sf

~~ \vspace{2cm}

\begin{center}

{\huge \sf
Iterative Solutions \\
for Low Lying Excited States  \\
\vspace{.3cm}
 of a Class of Schroedinger Equation$^*$}

\vspace{2cm}

{\large \sf

R. Friedberg$^1$, T. D. Lee$^{1,~2}$ and W. Q. Zhao$^{2,~3}$\\
} \vspace{.5cm}
{\it 1. Physics Department, Columbia University}\\
{\it New York, NY 10027, U.S.A.}\\
{\it 2. China Center of Advanced Science and Technology (CCAST/World Lab.)}\\
{\it P.O. Box 8730, Beijing 100080, China}\\
{\it 3. Institute of High Energy Physics, Chinese Academy of
Sciences\\
Beijing 100039, China  }

\end{center}

\vspace{0.5cm}
\begin{center}

{\large \sf Abstract}

\end{center}

{\sf The convergent iterative procedure for solving the
groundstate Schroedinger equation is extended to derive the
excitation energy and the wave function of the low-lying excited
states. The method is applied to the one-dimensional quartic
potential problem. The results show that the iterative solution
converges rapidly when the coupling $g$ is not too small. }

\vspace{.5cm}

{\normalsize \sf PACS{:~~11.10.Ef,~~03.65.Ge}}

\vspace{.5cm}


\vspace{1cm}

----------------------------------

\noindent * ~This research was supported in part by the U. S.
Department of Energy Grant

 DE-FG02-92ER-40699 and the
National Natural Science Foundation of China No.10547001}

\newpage

\begin{center}
 {\large \bf 1. Introduction}
\end{center}

{\large \sf

Consider the $N$-dimensional Schroedinger equation
$$
(-\frac{1}{2}{\bf \nabla}^2 +V(q)-E)\psi(q)=0 \eqno(1.1)
$$
with
$$
q =(q_1,~q_2,~\cdots,~q_N) \eqno(1.2)
$$
and
$$
{\bf \nabla}^2 = \sum_{i=1}^{N}\frac{\partial^2}{\partial q_i^2}.
$$
Through a linear transformation of the coordinates
$q_1,~q_2,~\cdots,~q_N$, (1.1) can be applied to most of the
nonrelativistic many body problems. Similarly, in the limit
$N\rightarrow \infty$, the same equation can also be extended to
relativistic bosonic field theories. However, except in a few
special cases, it is difficult to derive analytical solutions.
Recently, for a class of Schroedinger equations we have succeeded
in deriving a convergent iterative series solution for the
groundstate[1-4]. These include the Sombrero shaped potential in
any space dimension and arbitrary angular momentum[5]. In this
paper, we discuss possible extension to low lying excited states.

Let
$$
\psi_{gd}(q)=e^{-S(q)}\eqno(1.3)
$$
and $E_{gd}$ be the ground state wave function and the ground
state energy. Denote the corresponding ones for the first excited
state as
$$
\psi_{ex}=e^{-S}\chi~~~{\sf and}~~~E=E_{gd}+\epsilon.\eqno(1.4)
$$
We assume the potential $V(q)$ to have high barriers and low
valleys, so that the excitation energy $\epsilon$ is very small.
As an example, take $q$ to be a one-dimensional space $x$ and
$$
V(x)=\frac{g^{2}}{2} (x^{2}-1)^{2}     \eqno(1.5)
$$
with $g$ large. The first excited state would have an excitation
energy
$$
\epsilon \sim e^{-\frac{4}{3}g}<<1.\eqno(1.6)
$$
For clarity, we shall also assume that, in the general case of
$V(q)$, there is only one such low-lying state, as in the example
of the one-dimensional quartic potential.

Let the Schroedinger equations for $\psi_{gd}=e^{-S}$ and
$\psi_{ex}=e^{-S}\chi$ be
$$
(-\frac{1}{2}{\bf \nabla}^2 +V-E_{gd})e^{-S}=0 \eqno(1.7)
$$
and
$$
(-\frac{1}{2}{\bf \nabla}^2 +V-E_{gd})e^{-S}\chi=\epsilon
e^{-S}\chi. \eqno(1.8)
$$
Multiplying (1.7) by $e^{-S}\chi$ and (1.8) by $e^{-S}$, we obtain
from their difference,
$$
-\frac{1}{2}{\bf \nabla}\cdot (e^{-2S} {\bf \nabla} \chi)=\epsilon
e^{-2S}\chi. \eqno(1.9)
$$
Replace (1.9) by a series of iterative equations: for $n\geq 1$
$$
-\frac{1}{2}{\bf \nabla}\cdot (e^{-2S} {\bf \nabla}
\chi_n)=\epsilon_n e^{-2S}\chi_{n-1} \eqno(1.10)
$$
and when $n=0$,
$$
e^{-S}\chi_0=~~{\sf a~trial~function}~\psi_{trial}\eqno(1.11)
$$
which satisfies the orthogonality condition between $\psi_{gd}$
and $\psi_{trial}$; i.e.
$$
\int e^{-2S}\chi_0 d^Nq=0.\eqno(1.12)
$$
When $\chi_{n-1}$ is known, the iteration of (1.10) gives only the
ratio $\chi_n/\epsilon_n$. To perform the next order iteration it
is necessary to separate $\epsilon_n$ and $\chi_n$. For this
purpose, we may choose a fixed point $q^0$ and set
$$
\chi_n(q^0)=\chi_0(q^0).\eqno(1.13)
$$
Clearly, different choices of $q^0$ will yield different sequences
of $\chi_n(q)$ and $\epsilon_n$, as we shall discuss.

For given $e^{-S}$ and $\chi_{n-1}$, (1.10) can be viewed as an
analog problem in electrostatics with a dielectric constant
$\kappa$ and an external electric charge $\sigma_n$ given by
$$
\kappa=e^{-2S}~~~{\sf and}~~~\sigma_n=\epsilon_n
e^{-2S}\chi_{n-1}\eqno(1.14)
$$
Let $\vec{E}_n$ be the corresponding electrostatic field and
$\vec{D}_n$ the displacement field. We write
$$
\vec{E}_n=-\frac{1}{2}{\bf \nabla} \chi_n\eqno(1.15)
$$
and
$$
\vec{D}_n=\kappa\vec{E}_n.\eqno(1.16)
$$
Thus, (1.10) becomes
$$
{\bf \nabla}\cdot \vec{D}_n=\sigma_n.\eqno(1.17)
$$
At $\infty$, $\psi_{gd}=e^{-S}=0$. Therefore
$\kappa(\infty)=0,~\sigma_n(\infty)=0$ and
$$
\vec{D}_n(\infty)=0.\eqno(1.18)
$$
Integrating (1.17) over all space we find, on account of (1.18),
$$
\int\sigma_nd^Nq=\int e^{-2S}\chi_nd^Nq=0.\eqno(1.19)
$$
Hence, the orthogonality condition (1.12) is now carried over to
all $n\geq 1$.

Expand $e^{-S}\chi _{n}$ in terms of the set of all eigenstates
$\{\psi_{a}\}$ of (1.1). We have, because of (1.19),
$$
e^{-S}  \chi _{n}=\sum\limits_{a\neq gd} c_{a} (n)\psi_{a}
\eqno(1.20)
$$
with
$$
c_{a}(n)=\frac{1}{E_{a}-E_{gd}} \epsilon_{n}\cdot c_{a}
(n-1).\eqno(1.21)
$$
For problems like the one-dimensional quartic potential (1.5),
when the coupling $g^2$ is large, only the first excited state
$a=1$ has an excitation energy $\epsilon=E_1-E_{gd}$ that is
exponentially smaller than all other $E_{a}-E_{gd}$. Thus, the
corresponding $c_1(n)$ becomes exponentially large compared with
all other $c_{a\neq 1}(n)$, in accordance with (1.21). Hence the
iteration process (1.10) becomes rapidly convergent.

For a large class of problems in which $V(q)$ has high barriers
and low valleys, there are often only a finite number of low-lying
excited states.
$$
a=1,~2,~\cdots,~m\eqno(1.22)
$$
with excitation energies $\epsilon_a=E_a-E_{gd}$ comparable to
each other, but exponentially smaller than those of $a>m$ states.
By maintaining the orthogonality relations between these $m$
low-lying excited states, the iteration process (1.10) can be
readily generalized to such problems.


\begin{center}
 {\large \bf 2. One Dimensional Problem}
\end{center}

In one dimension, we replace $\{q_i\}$ by a single $x$. For
simplicity, consider the special case
$$
V(x)=V(-x).\eqno(2.1)
$$
Thus, the groundstate wave function $e^{-S(x)}$ is an even
function of $x$ and the first excited state $e^{-S(x)}\chi(x)$ an
odd function. We assume that $e^{-S(x)}$ is already known, e.g. by
following the method discussed in Refs.[1-4]. We further assume
that the first excited state has an excitation energy
$$
\epsilon<<1,\eqno(2.2)
$$
as would be the case if $V(x)$ is like the quartic potential (1.5)
with large $g$. In one-dimension, (1.9) becomes
$$
-\frac{1}{2}\frac{d}{dx} (e^{-2S} \frac{d\chi}{dx})=\epsilon
e^{-2S}\chi. \eqno(2.3)
$$
The corresponding series of iterative equations (1.10) for $n\geq
1$ is
$$
-\frac{1}{2}\frac{d}{dx} (e^{-2S} \frac{d\chi_n}{dx})=\epsilon_n
e^{-2S}\chi_{n-1} \eqno(2.4)
$$
with $\chi_0$ a properly chosen trial function. As in (1.13), in
order to separate $\chi_n$ and $\epsilon_n$ from the ratio
$\chi_n/\epsilon_n$ we choose a fixed point $x^0$ and set
$$
\chi_n(x^0)=\chi_0(x^0)\eqno(2.5)
$$
for all $n$.

In terms of the electrostatic analog (1.14)-(1.16), write
$$
\kappa=e^{-2S},~~~~~\sigma_n=\epsilon_ne^{-2S}\chi_{n-1}
$$
$$
E_n=-\frac{1}{2}\chi_n'~~~~{\sf and}~~~~D_n=\kappa E_n.\eqno(2.6)
$$
with $'$ denoting $d/dx$. Thus, (1.17) and (1.18) become
$$
D_n'(x)=\sigma_n(x)\eqno(2.7)
$$
$$
D_n(\pm\infty)=0~~{\sf and}~~\sigma_n(\pm\infty)=0.\eqno(2.8)
$$
From (2.7) and (2.8), we have
$$
D_n(x)=-\epsilon_n\int\limits_x^\infty
e^{-2S(z)}\chi_{n-1}(z)dz,\eqno(2.9)
$$
and therefore
$$
E_n(x)=e^{2S(x)}D_n(x)\eqno(2.10)
$$
can also be readily expressed in terms of $\chi_{n-1}$. Since
$\chi_n(x)$ is odd in $x$, we need only to consider $x\geq 0$.
From (2.6) and (2.10), we find
$$
\chi_n(x)=2\epsilon_n\int\limits_0^x e^{2S(y)}dy
\int\limits_y^\infty e^{-2S(z)}\chi_{n-1}(z)dz.\eqno(2.11)
$$
Since $e^{-S(x)}$ is the groundstate, it has no zero at finite
$x$. Thus, the factor $e^{2S(y)}$ in the $y$-integration of (2.11)
is always finite. This is an important fact that enables us to
extend the effectiveness of the iterative procedures of Refs.[1-5]
for the groundstate to the low-lying excited state.

\begin{center}
 {\large \bf 3. An Analytically Soluble Example}
\end{center}

As an analytically soluble example, we consider the following
simple one dimensional example
\begin{eqnarray*}\label{3.1}
~~~~~~~~~~~~~~~~~~~~~~~~~~V(x)=\lambda\delta(x)+
 \left\{
\begin{array}{ll}
0~~,~~~~~~~~|x|<1\\
\infty~,~~~~~~~~|x|>1.
\end{array}
\right.~~~~~~~~~~~~~~~(3.1)
\end{eqnarray*}
The unnormalized groundstate solution is
\begin{eqnarray*}\label{3.2}
~~~~~~~~~~~~~~~e^{-S(x)}=
 \left\{
\begin{array}{ll}
\sin p(1-x)~~~~~~~~~~~~~~~0<x<1\\
~~~~~~0~~~~~~~~~~~~~~{\sf for}~~~~~~~~|x|>1 \\
\sin p(1+x)~~~~~~~~~~~~~-1<x<0
\end{array}
\right.~~~~~~~~~~~~~~~(3.2)
\end{eqnarray*}
with
$$
p\equiv \pi-\delta
$$
$$
{\sf and}~~~~~~~~~~~~~~~~~~~~~~~~~~~~~~~~~~~~~~~~~~~~~~
~~~~~~~~~~~~~~~~~~~~~~~~~~~~~~~~~~~~~~~\eqno(3.3)
$$
$$
\lambda=-p\cot p =(\pi-\delta)\cot \delta.
$$
For very large $\lambda$, $p$ is very close to $\pi$, so
$$
\delta \ll 1.\eqno(3.4)
$$
The exact lowest excited state is, for $|x|<1$
$$
e^{-S}\chi =\sin\pi x  \eqno(3.5)
$$
with the excitation energy
$$
\epsilon=\frac{1}{2} (\pi^{2}-p^{2}) =\pi\delta- \frac{1}{2}
\delta^{2} \ll 1, \eqno(3.6)
$$
and for $x\geq 0$, $\chi$ is given by
$$
\chi (x)=\frac{\sin\pi x}{\sin p(1-x)}.    \eqno(3.7)
$$
In order to test the effectiveness of the iterative solution
(2.11) and the supplementary condition (2.5), we shall start from
the groundstate solution (3.2) and a trial function
$$
\chi_0=x.   \eqno(3.8)
$$
Since (2.11) yields only the ratio $\chi_n(x)/\epsilon_n$, we
follow (2.5) by choosing
$$
x^0=1\eqno(3.9)
$$
and therefore
$$
\chi_n(1)=\chi_0(1)=1.\eqno(3.10)
$$
It is clear that the trial function
$$
e^{-S}\chi_0 =x\sin p(1-x)\eqno(3.11)
$$
is \underline{not} a very good guess of the first excited state
(3.5), nor does (3.8) resembles (3.7), except that at $x=0$,
$\chi_0(0)=\chi(0)=0$. Nevertheless, we shall show that the
iterative solution (2.11) with the supplementary condition (3.10)
does lead to a rapidly convergent sequence.

Substituting (3.2) and (3.8) into (2.11), we find, for $x\geq 0$,
$$
[\frac{4p}{2\epsilon_{1}}\sin p]  \chi_1 e^{-S} =\sin px-(\sin
p)[\frac{x}{p}\sin p(1-x)+x^{2}\cos p(1-x)].\eqno(3.12)
$$
The supplementary condition (3.10) leads to
$$
\epsilon_1=\frac{2p^2}{1-p\cot p}.\eqno(3.13)
$$
For $\delta=\pi-p$ small, (3.13) gives
$$
\epsilon_1=2\pi\delta-4\delta^2+2(\frac{1}{\pi}
+\frac{\pi}{3})\delta^3-2\delta^4+O(\delta^5) \eqno(3.14)
$$
whereas the exact $\epsilon$ is given by (3.6) with
$\epsilon_1\cong 2\epsilon$. However, on second and third
iterations, we find
$$
\epsilon_2=\pi\delta+(\frac{1}{\pi}-\frac{\pi}{3})\delta^3
+(\frac{1}{3}-\frac{2}{\pi^2})\delta^4+O(\delta^5)\eqno(3.15)
$$
$$
\epsilon_3=\pi\delta-\frac{1}{2}~\delta^2+\frac{-6+\pi^2}{12\pi}~\delta^3
-\frac{15}{8\pi^2}~\delta^4+O(\delta^5).\eqno(3.16)
$$
Comparing to the exact value, the second order solution
$\epsilon_2$ gives the correct $\pi \delta$ and the third order
$\epsilon_3$ gives the correct first two terms: $\pi \delta
-\frac{1}{2}~\delta^2$.

For $\delta=0.1$, (3.14)-(3.16) yield
$$
\epsilon_1=0.59086,~~\epsilon_2=0.31348,~~\epsilon_3=0.30924\eqno(3.17)
$$
which can be compared with the exact solution
$$
\epsilon=0.30916.
$$
The $\chi_n(x)$ for $n=0,~1,~2,~3$ are plotted in Fig.1, together
with the exact solution $\chi(x)$. In fact, $\chi_2$, $\chi_3$ and
$\chi$ are on the same curve. This shows the rapid convergence of
the iterative process.

\newpage

\begin{figure}[h]
 \centerline{
\epsfig{file=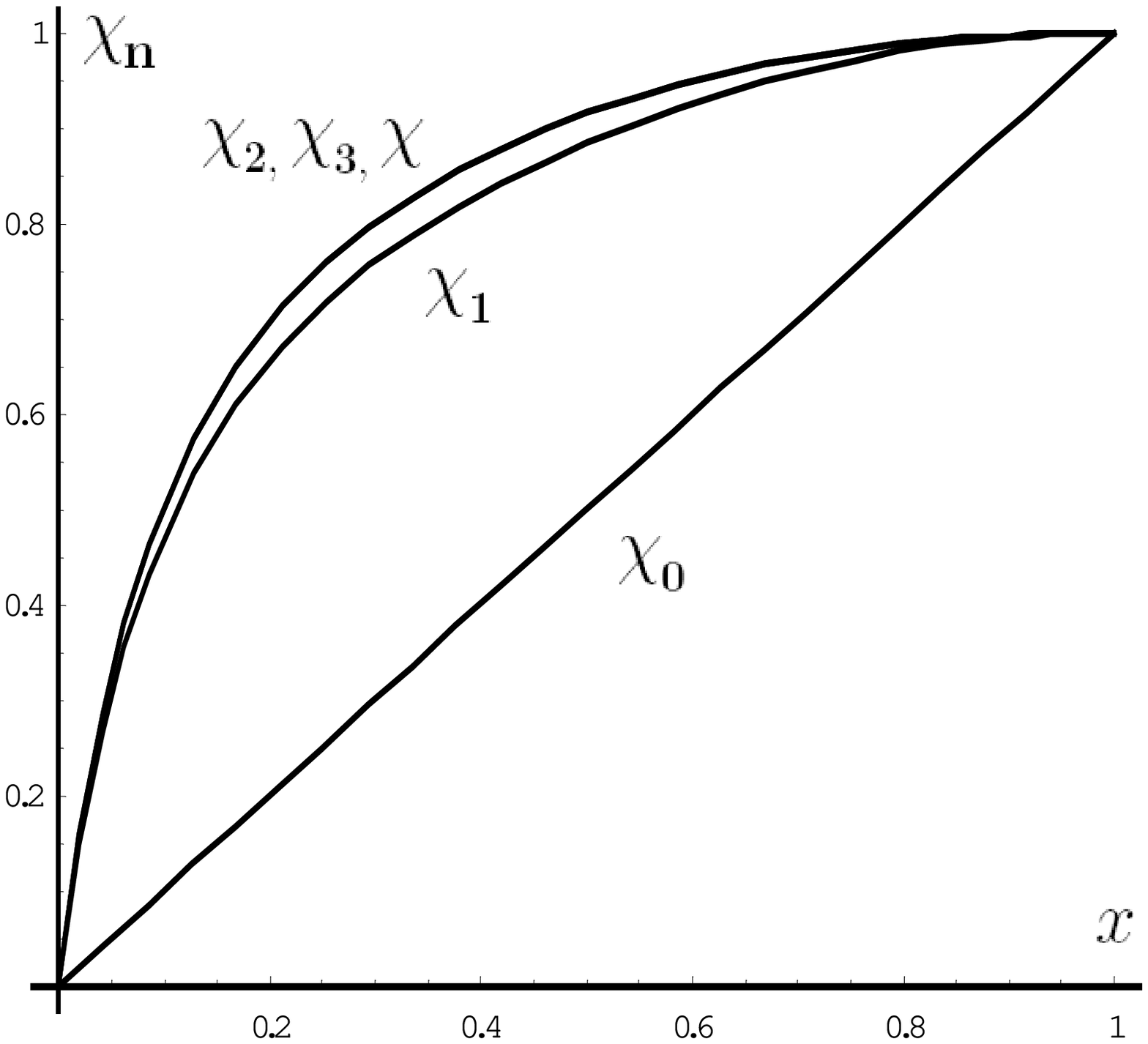, width=12cm, height=11cm}} \vspace{.2cm}
 \centerline{{\normalsize \sf Fig.1~ $\chi$ and $\chi_n$ for $n=0,~1,~2,~3$ for the soluble problem (Section 3). }}
\end{figure}

\newpage

\begin{center}
 {\large \bf 4. One-dimensional Quartic Potential}
\end{center}

Consider the one-dimensional quartic potential
$$
V=\frac{g^{2}}{2} (x^{2}-1)^{2}        \eqno(4.1)
$$
Assume that the groundstate wave function $e^{-S}$ has been
obtained by using the method described in our recent paper[5]. The
first excited state $e^{-S}\chi$ satisfies (2.3) and the
corresponding $n^{\sf th}$ order iterative solution is
$e^{-S}\chi_n$, with $\chi_n$ given by (2.11). When $n=0$, we
choose the trial function $\chi_0$ to be an odd function, with
$$
\chi_0(-x)=-\chi_0(x)
$$
and for $x$ positive
\begin{eqnarray*}\label{4.2}
~~~~~~~~~~~~~~~~\chi_0(x)=
 \left\{
\begin{array}{ll}
x(2-x)~~~~~{\sf for}~~~~0<x<1\\
~~~~~~1~~~~~~~~~{\sf for}~~~~~~~~x>1
\end{array}
\right.~~~~~~~~~~~~~~~~~~~~~~(4.2)
\end{eqnarray*}
As in (2.5) and (3.9)-(3.10), we choose $x^0=1$ and set
$$
 \chi_{n}(1)= \chi_{0}(1)=1.\eqno(4.3)
$$
In the following, we also set
$$
g=3.\eqno(4.4)
$$
Using the iteration method of Refs.[1,5], we find the groundstate
energy to be
$$
E_{gd}=2.48291.\eqno(4.5)
$$
The lowest excitation energy $\epsilon_n$ for the first four
iterations ($n=1,~2,~3,~4$) based on (2.11) and (4.3) are
$$
\epsilon_1=0.41776,~\epsilon_2=0.41367,
$$
$$
\epsilon_3=0.413568,~\epsilon_4=0.413568.\eqno(4.6)
$$
Thus, with three iterations, $\epsilon_3$ is already accurate to
seven significant figures. The corresponding wave functions
$\chi_n$ are plotted in Figure 2. The groundstate $e^{-S}$ and the
first excited state $e^{-S}\chi$ are given in Figure 3. To the
same seven significant figures the eigenvalue for the first
excited state can be expressed as $E_{odd}=E_{gd}+\epsilon_4$.

To test the sensitivity to the choice $x^0$ we set, instead of
(4.3), $x^0=1/2$ and require
$$
\chi_n(\frac{1}{2})=\chi_0(\frac{1}{2}).\eqno(4.7)
$$
The corresponding excitation energies $\epsilon_n$ of the lowest
excited state for the first 4 iterations are, in place of (4.6),
$$
\epsilon_1=0.41363,~\epsilon_2=0.41358
$$
$$
\epsilon_3=0.413569,~\epsilon_4=0.413568.\eqno(4.8)
$$
Although the value of $\epsilon_1$ in (4.8) differs from the
corresponding one in (4.6), after the 4th iteration $\epsilon_4$
becomes essentially the same for both cases. This means that the
iteration result is not too sensitive to the choice of $x^0$.

\begin{figure}[h]
 \centerline{
\epsfig{file=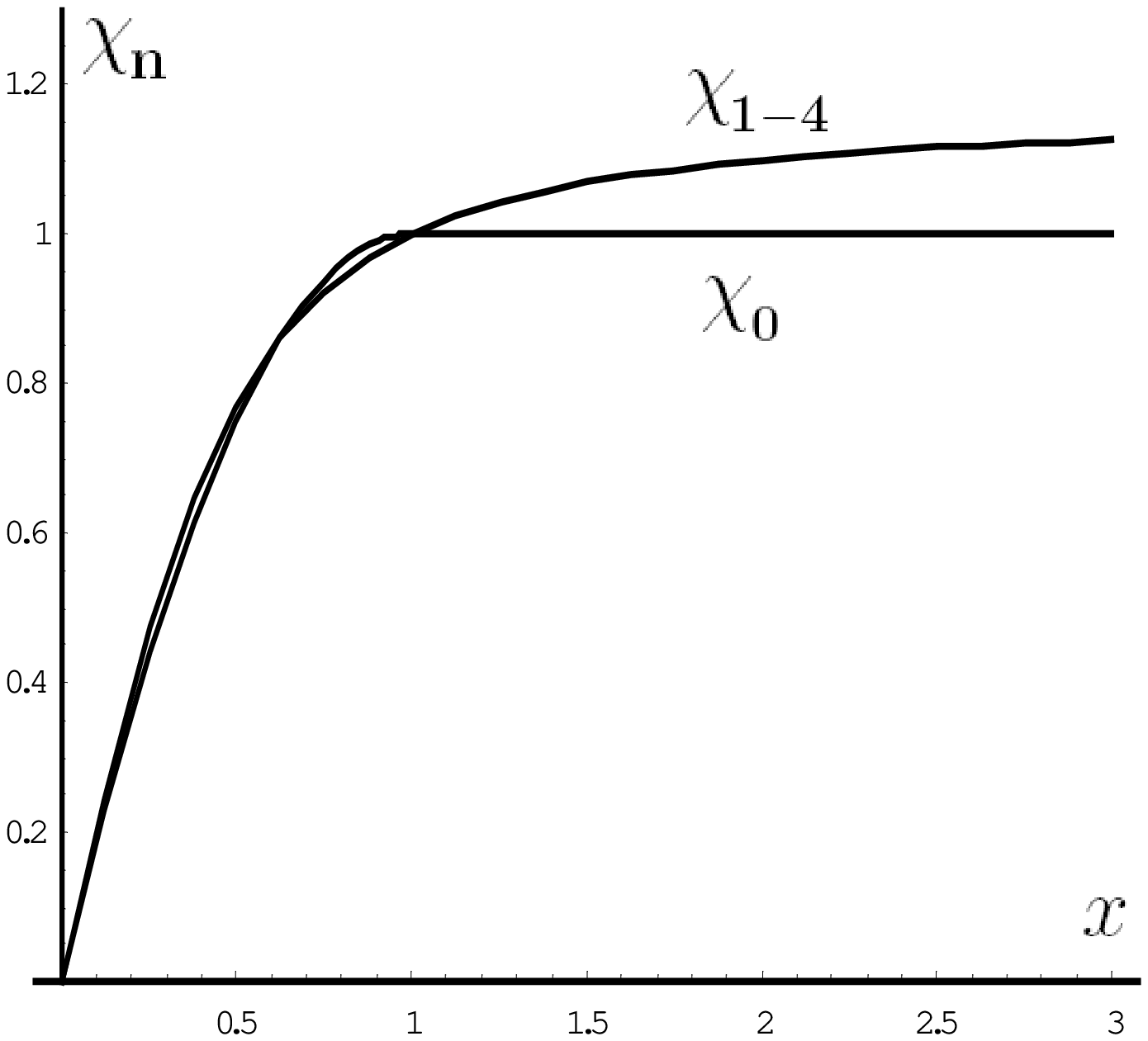, width=12cm, height=11cm}} \vspace{.2cm}
 \centerline{{\normalsize \sf Fig.2~ $\chi_n$ for $n=0,~1,~2,~3,~4$ for the quartic potential (4.1) with $g=3$. }}
\end{figure}

\newpage

\begin{figure}[h]
 \centerline{
\epsfig{file=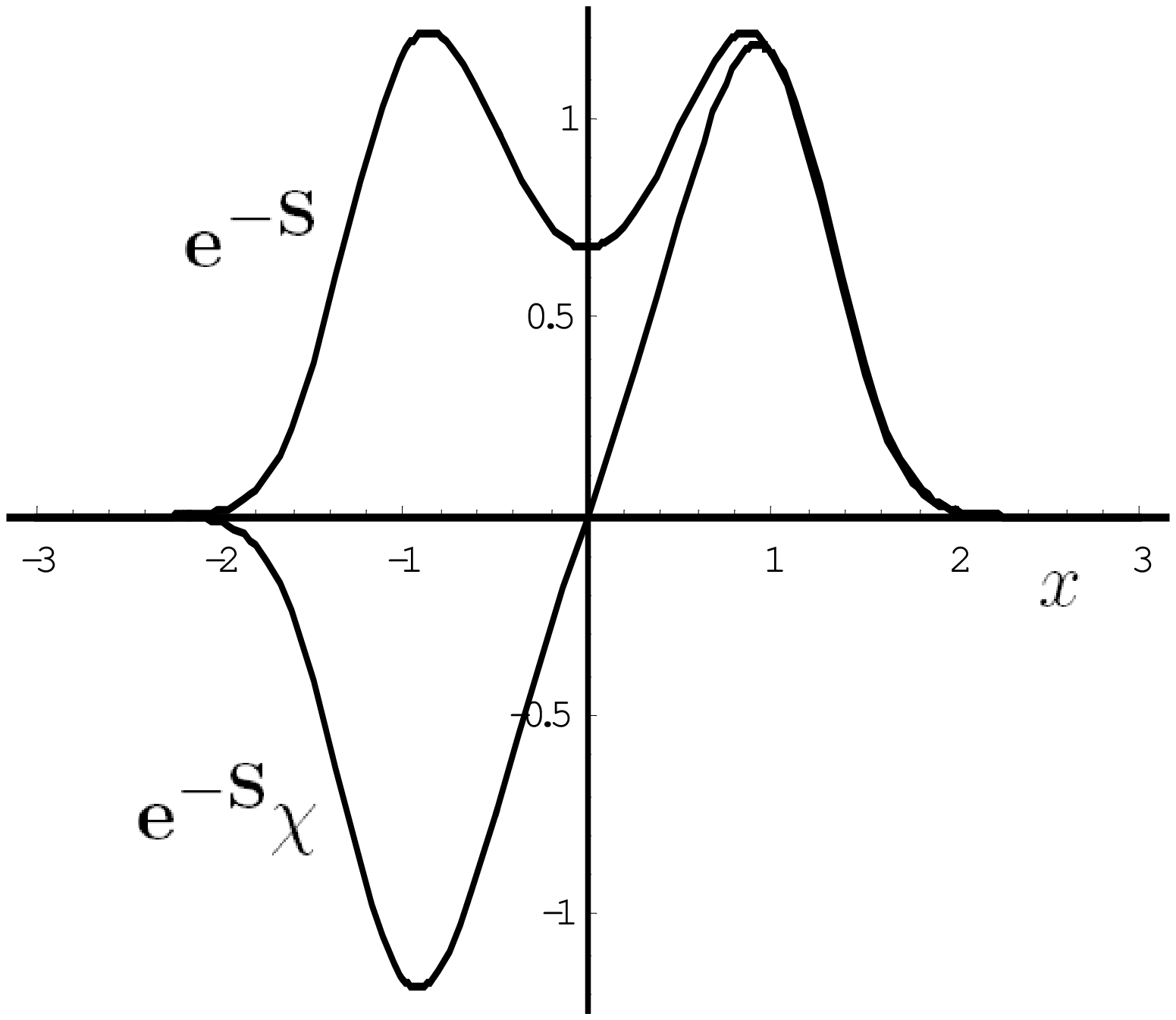, width=12cm, height=11cm}} \vspace{.2cm}
 \centerline{{\normalsize \sf Fig.3~ The groundstate wave function $e^{-S}$
 and the first excited state }}
  \centerline{{\normalsize \sf ~~~~~~~wave function $e^{-S}\chi$ for the
quartic potential (4.1) with $g=3$. }}
\end{figure}

\newpage

\begin{center}
 {\large \bf Appendix}
\end{center}

In our earlier work[6] an asymptotic expansion of the average $E=
\frac{1}{2}(E_{gd}+E_{odd})$ and that of the difference
$\Delta=\frac{1}{2}~\epsilon$ were obtained. To compare our
present results with these asymptotic expansions, we take a larger
coupling $g=8$. This enables us to use the asymptotic expansion up
to order $1/g^3$ for the wave function $\phi_\pm$ of Ref.[6]. We
find
$$
E_{asymp}=7.728854~.\eqno(A.1)
$$
Using the expressions
$$
\Delta =\frac{\lambda}{2}~\frac{1}{\int\limits_0^\infty
\phi^2_+(x)dx}\eqno(A.2)
$$
and
$$
\lambda=\phi'_+\phi_--\phi_+\phi'_-\eqno(A.3)
$$
of Ref.[6], we obtain
$$
\epsilon_{asymp}=2\Delta_{asymp}=0.003027.\eqno(A.4)
$$
For $g=8$, the iterative method of Ref.[5] gives
$$
E_{gd}= 7.727340.\eqno(A.5)
$$
The corresponding $n^{{\sf th}}$ order iterative excitation energy
$\epsilon_n$ of Sec. 4 for $n=1-4$ are
$$
\epsilon_1=0.00310125,~\epsilon_2=0.00301796
$$
$$
\epsilon_3=0.003017947,~\epsilon_4=0.003017947.\eqno(A.6)
$$
Keeping the accuracy to seven significant figures, we find
$$
E=E_{gd}+\frac{1}{2}~\epsilon=7.728849\eqno(A.7)
$$
and
$$
\epsilon=0.003018.\eqno(A.8)
$$
Thus, the asymptotic values $E_{asymp}$ and $\epsilon_{asymp}$ of
(A.1) and (A.4) compare favorably well with $E$ and $\epsilon$ of
(A.7) and (A.8). However, inclusions of still higher and higher
order terms in the asymptotic expansion would lead to divergent
results for $E_{asymp}$ and $\epsilon_{asymp}$.

The wave functions $\chi_n(x)$ for $g=8$ are plotted in Fig. 4. As
we can see, the deviation  of the $\chi_n(x)$ from $1$ happens
only at small $x$.


\begin{figure}[h]
 \centerline{
\epsfig{file=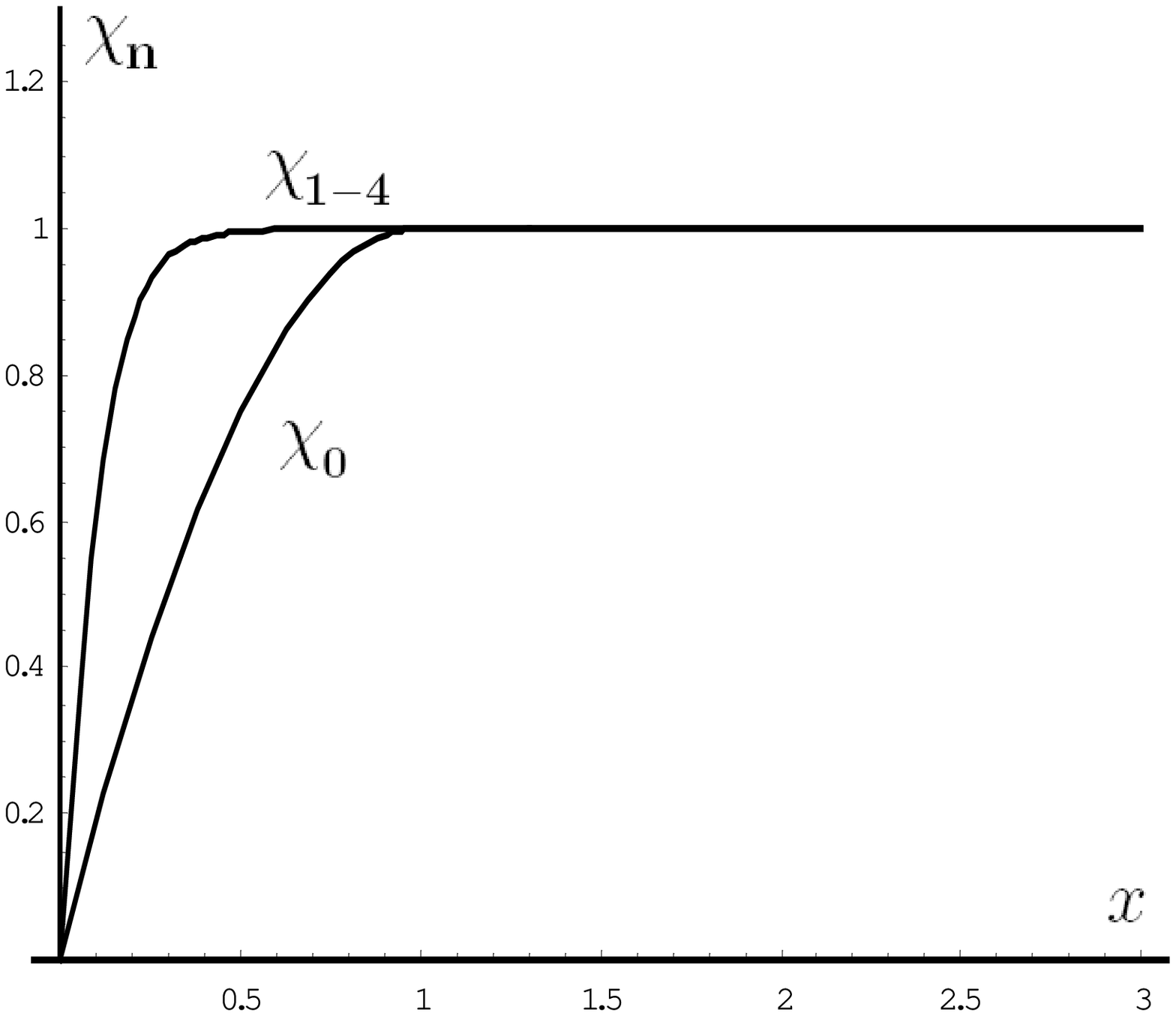, width=12cm, height=11cm}} \vspace{.2cm}
 \centerline{{\normalsize \sf Fig.4~ $\chi_n$ for $n=0,~1,~2,~3,~4$ for the quartic potential (4.1) with $g=8$. }}
\end{figure}


\begin{center}
 {\large \bf References}
\end{center}

}
{\sf

 [1] R. Friedberg, T. D. Lee, W. Q. Zhao and A. Cimenser, Ann.
Phys. 294(2001)67

[2] R. Friedberg and T. D. Lee, Ann. Phys. 308(2003)263

[3] R. Friedberg and T. D. Lee, Ann. Phys. 316(2005)44

[4] T. D. Lee, J. of Stat. Phys. 121(2005)1015

[5] R. Friedberg, T. D. Lee and W. Q. Zhao, Ann. Phys. in press,
quant-ph/0510193

[6] R. Friedberg, T. D. Lee and W. Q. Zhao, IL Nuovo Cimento
112A(1999)1195

}

\end{document}